\documentclass[prd,onecolumn,showpacs,floatfix,superscriptaddress,nofootinbib]{revtex4}
\usepackage{graphicx}
\usepackage{epsfig}
\usepackage{bm}
\usepackage{amssymb}
\usepackage{float}
\usepackage{amsmath}
\usepackage{dcolumn}
\usepackage{cancel}
\usepackage[colorlinks]{hyperref}
\usepackage[usenames,dvipsnames]{color}
\hypersetup{
     breaklinks=true,
    pdfstartview={FitH},    
    colorlinks=true,       
    linkcolor=blue,          
    citecolor=red,        
    filecolor=magenta,      
    urlcolor=blue,           
    anchorcolor=green,      
    linktocpage=true
}

\renewcommand{\k}{\vec{k}}

\def\doi{http://doi.org}

\def\r{\mathrm{r}}
\def\g{\mathrm{g}}

\def\m{\mathrm{m}}

\def\s{\mathrm{s}}

\def\Lm{ \mathcal{L}_m}
\def\a{ \alpha}
\def\b{ \beta}
\def\g{ \gamma}
\def\d{ \delta}
\def\m{ \mu}
\def\n{ \nu}

\def\r{ \rho}
\def\s{ \sigma}
\def\k{ \kappa}

\def\hd{\dot{H}}

\def\be{\begin{equation*}}
\def\ee{\end{equation*}}

\def\a{ \alpha}
\def\b{ \beta}
\def\g{ \gamma}
\def\d{ \delta}
\def\m{ \mu}
\def\n{ \nu}

\def\r{ \rho}
\def\s{ \sigma}
\def\k{ \kappa}

\begin{document}

\title{Cosmology in cubic and $f(P)$ gravity}

\author{Cristian Erices}\email{crerices@central.ntua.gr}
\affiliation{Department of Physics, National Technical University of Athens, Zografou
Campus GR 157
73, Athens, Greece}
\affiliation{ Universidad Cat\'olica del Maule, Av. San Miguel 3605, Talca, Chile}

 \author{Eleftherios Papantonopoulos}
 \email{lpapa@central.ntua.gr}
  \affiliation{Department of Physics, National Technical University of Athens, Zografou
Campus GR 157
73, Athens, Greece}

\author{Emmanuel N. Saridakis}
\email{msaridak@phys.uoa.gr}
\affiliation{Department of Physics, National Technical University of Athens, Zografou
Campus GR 157 73, Athens, Greece}
\affiliation{Department of Astronomy, School of Physical Sciences, University of Science
and Technology of China, Hefei 230026, China}


\begin{abstract}
We construct cubic gravity and its $f(P)$ extension and we investigate their early- and 
late-time cosmological applications. Cubic gravity is based on
a particular invariant $P$, constructed from cubic contractions of the
Riemann tensor, under three requirements: (i)  the resulting theory possesses a
spectrum identical to that of general relativity, (ii) it is neither topological nor
trivial in four dimensions, and (iii) it is defined such that it is independent of the 
dimensions. Relaxing the last condition and restricting the parameters of cubic gravity we 
can obtain second-order field equations in a cosmological background.
We show that at early times one can obtain inflationary, de Sitter solutions,
which are  driven by an effective cosmological constant constructed purely from
the cubic terms of the simple cubic or $f(P)$ gravity.
Concerning late-time evolution, the new terms constitute an effective dark-energy 
sector and we show 
that the Universe experiences the usual thermal history and the onset  of late-time 
acceleration. In the case of $f(P)$ gravity, depending on the choice of parameters, we 
find that the dark-energy equation-of-state parameter can be quintessencelike, 
phantomlike or it can experience the phantom-divide crossing during the evolution, even if an 
explicit cosmological constant is absent.
\end{abstract}

\maketitle

\section{Introduction}

Higher-order gravities have been introduced in the general framework of modified theories 
of gravity, with the aim to describe in a uniform way the history of the Universe; to 
account for the early-time inflation, the late-time acceleration, and the presence of dark matter; 
and to be consistent with observations 
\cite{Nojiri:2006ri,Copeland:2006wr,Nojiri:2010wj,Clifton:2011jh,Cai:2015emx}. A more 
theoretical motivation in studying  higher-order corrections to the Einstein-Hilbert term 
is that these theories arise naturally in the gravitational effective action of a complete 
string theory \cite{Gross:1986mw} and they result in a renormalizable and thus quantizable 
gravitational theory \cite{Stelle:1976gc}.  Additionally,  in such considerations one can 
construct theories which possess general relativity as a
particular limit \cite{Biswas:2011ar}. Moreover, certain higher-order gravities  are 
equivalent to Einstein gravity at the linearized level in the vacuum, and the only 
physical mode propagated by 
the metric perturbation is a transverse and massless graviton. Such theories are certain  
f$($Lovelock$)$ theories \cite{Lovelock:1971yv}.

Higher-order gravities and more generally modified theories of gravity provide a deeper 
understanding of Einstein gravity theory itself.
The construction of modified theories of gravity starts
from the Einstein-Hilbert Lagrangian and includes extra terms, such as in $f(R)$ gravity 
\cite{Starobinsky:1980te,Capozziello:2002rd,DeFelice:2010aj,Nojiri:2010wj}, $f(G)$
gravity \cite{Nojiri:2005jg, DeFelice:2008wz}, Lovelock
gravity \cite{Lovelock:1971yv, Deruelle:1989fj}, Weyl gravity
\cite{Mannheim:1988dj, Flanagan:2006ra}, Galileon theory
\cite{Nicolis:2008in, Deffayet:2009wt, 
Leon:2012mt,Kolyvaris:2011fk,CisternaErices,Erices:2015xua,
Erices:2017izj}, etc.
More radical modifications of Einstein gravity theory are provided by the introduction of 
torsion terms in $f(T)$ gravity
\cite{Ben09, Linder:2010py, Chen:2010va,Kofinas:2015hla} or $f(T,T_G)$ gravity
\cite{Kofinas:2014owa,Kofinas:2014daa}. These modified theories of gravity allow us to 
unveil what features of a gravitational theory are generic and which are specific.

A potential disadvantage of the curvature-based corrections to the Einstein-Hilbert
action is that the couplings of the different curvature invariants depend on the 
spacetime dimension $D$. Hence, they are actually different theories in different 
dimensions. Furthermore, the higher-order 
terms may lead to the appearance of higher-than-second-order derivatives in the field equations. Although this does not necessarily mean that
the theory presents instabilities and pathologies, since the higher-order derivatives may
be just a reflection of healthy extra degrees of freedom (this can be shown through a
Hamiltonian analysis), keeping the field equations up to second order is a desirable
feature since it does ensure that the theory is pathology free.

One interesting class of modified gravity based on higher-order curvature invariants was
recently constructed in \cite{Bueno:2016xff}, and it uses cubic contractions of the
Riemann tensor. This theory, called Einsteinian cubic gravity (ECG), possesses   
basic health conditions coming from nontopological terms. Namely, the coupling 
parameters, which are dimension independent, allow to propagate a massless and transverse 
graviton on a maximally symmetric background in the same way as the standard general 
relativity (GR). In general, such terms contribute with fourth-order derivatives of the 
metric in the field equations. However, as it was 
shown in \cite{BuenoCano,Mann01,Mann02}, the original form of the theory is sufficient to 
admit 
spherically symmetric black hole solutions with a second-order differential equation for the 
metric 
function. An extra cubic correction, which is trivial for a spherically symmetric black 
hole ansatz, 
allows us additionally to accommodate a Friedmann-Lema\^{i}tre-Robertson-Walker (FLRW) 
solution with 
second-order field equations for the scale factor, leading to a purely geometric 
inflationary 
period \cite{Edelstein01} \footnote{Black holes solutions were also obtained in 
\cite{Edelstein02,CisternaQTG}, alongside the examination of a well-defined cosmology 
for higher order terms such as the quartic, quintic and  infinite tower of 
higher-curvature corrections to the Einstein-Hilbert action. }. All these features make 
this theory physically interesting and hence worthy of 
further 
investigation.

In this work we  study the cosmological applications of cubic gravity, as well as investigate its modifications, namely $f(P)$ gravity. In particular, we are 
interested in examining whether the cubic terms $P$ can drive the Universe's acceleration at 
early and late times, and whether we can obtain the usual thermal history of the Universe. 
We show that since we are considering four-dimensional spacetimes, the condition of 
dimension-independent coupling parameter can be relaxed,\footnote{This can be achieved by 
considering the dimension-dependent parameters of ECG 
in \cite{Bueno:2016xff}, namely considering the expressions for the parameters before 
the dimension-independence condition is imposed, and evaluating in four dimensions.} 
allowing us to tune the parameters in order to get second-order field equations for the 
scale factor.

Inspired by ECG, we obtain a modified theory that shares all the good properties of cubic 
gravity  in four dimensions. In particular, we consider an action  with corrections to the 
Einstein-Hilbert one, constituted by an arbitrary $f(P)$ function of the cubic term. This 
provides a remarkable advantage over the cosmological models studied in ECG, since it 
generates inflationary, de Sitter solutions at early times and reproduces the onset of 
late-time acceleration and the evolution of the densities in agreement with observations, 
even when the cosmological constant is zero.

The plan of the work is
as follows. In Sec. \ref{TheModel} we construct cubic gravity and we extend it to
the $f(P)$ case. Additionally, we apply these theories in a cosmological framework and we
extract the Friedmann equations. In Sec. \ref{CosmAppl} we perform a detailed
investigation of the Universe's behavior at both early and late times, examining various
observables such as the effective dark-energy density and equation-of-state
parameters. Finally, in Sec. \ref{Conclusions}
 are our conclusions.

\section{Cubic and $f(P)$ gravity}
\label{TheModel}

In this section we present cubic gravity, as well as its $f(P)$ extension, and we apply
them in a cosmological framework. The starting point of cubic gravity is to include
additional terms in the Einstein-Hilbert action that are constructed by contractions of
three Riemann tensors
\cite{Bueno:2016xff}. Nevertheless, one desires combinations that lead to a theory
with the following features:
(i) it possesses a spectrum identical to that of
general relativity, i.e. the metric perturbation on a maximally symmetric background
propagates only a transverse and massless graviton, and (ii) it is neither topological nor trivial in four dimensions. Additionally, we desire
that (iii) the theory leads to second-order field equations. In ECG 
an additional condition  is required, namely that the theory should be
constructed in a dimension-independent way. However, since in the present work we are
interested in the usual four-dimensional cosmological evolution, we relax such a 
requirement. This freedom allows us to tune the parameters to satisfy condition (iii) 
without the inclusion of additional cubic terms. 

Let us focus on four spacetime dimensions. In this case, a general 
nontopological [thus
condition (iii)   is satisfied] cubic term $P$  would be
\begin{eqnarray}\label{P}
&&P=\b_1
{{{R_{\m}}^{\r}}_{\n}}^{\s}{{{R_{\r}}^{\g}}_{\s}}^{\d}{{{R_{\g}}^{\m}}_{\d}}^{\n}+\b_2
R_{\m\n}^{\r\s}R_{\r\s}^{\g\d}R^{\m\n}_{\g\d}+\b_3
R^{\s\g}R_{\m\n\r\s}{R^{\m\n\r}}_{\g}+\b_4
R R_{\m\n\r\s}R^{\m\n\r\s}\nonumber\\
&&
\ \ \ \ \ \ 
+\b_5 R_{\m\n\r\s}R^{\m\r}R^{\n\s}+\b_6 R^{\n}_{\m}R^{\r}_{\n}R^{\m}_{\r}+\b_7
R_{\m\n}R^{\m\n}R+\b_8 R^3,
\end{eqnarray}
where   $\beta_i$ are parameters. One can show that condition (i) above  is satisfied
if we impose the parameter relations \cite{Bueno:2016xff}
\begin{eqnarray}
 \label{prop}
&&\b_7=\frac{1}{12}(3\b_1-24\b_2-16\b_3-48\b_4-5\b_5-9\b_6),\\
&&\b_8=\frac{1}{72}(-6\b_1+36\b_2+22\b_3+64\b_4+3\b_5+9\b_6).
\end{eqnarray}

One can now use $P$ as a correction term in  the Einstein-Hilbert action, and construct
the cubic gravity as
\begin{equation}
S=\int\sqrt{-g}d^4x\left[\frac{1}{2\kappa}(R-2\Lambda)+\a
P\right],
\label{action0}
\end{equation}
with $\alpha$ a coupling parameter, $\kappa=8\pi G$ the Newton's constant, and where we
have also allowed for an explicit cosmological constant $\Lambda$ for completeness. We
mention here that in principle one should include also quadratic corrections to the above
action, such as the Gauss-Bonnet combination
${\cal{G}}=R^2-R_{\m\n}R^{\m\n}+R_{\m\n\r\s}R^{\m\n\r\s}$. However, in this work we
focus on four dimensions and such a term can be neglected since it is topological.

Variation of the action $S+S_m$, where $S_m$ accounts for the matter Lagrangian $\Lm$,
leads
to the field equations
\begin{equation}\label{eom}
G_{\mu\nu}+\Lambda g_{\mu\nu}=\kappa (T_{\m\n}+\a H_{\m\n}),
\end{equation}
with
\begin{equation}
T_{\mu\nu}=-\frac{2}{\sqrt{-g}}\frac{\delta (\sqrt{-g}\Lm)}{\delta
g^{\mu\nu}},
\label{Tmn}
\end{equation}
representing the matter energy-momentum tensor and\footnote{We use a normalized symmetrization $A_{(\mu\nu)}:=\frac{1}{2}(A_{\mu\nu}+A_{\nu\mu})$ and antisymmetrization $A_{[\mu\nu]}:=\frac{1}{2}(A_{\mu\nu}-A_{\nu\mu})$.}
\begin{equation}\label{Hmn0}\\
\begin{aligned}
H_{\m\n}&=-\frac{2}{\sqrt{-g}}\frac{\delta (\sqrt{-g}P)}{\delta g^{\mu\nu}}\\
&=g_{\mu\nu}P+{R^{\a\b\r}}_{(\mu}K_{\nu)\r\a\b}+2\nabla^{\a}\nabla^{\b}K_{\a(\m\n)\b},
\end{aligned}
\end{equation}
where  $\nabla_{\mu}$ is the covariant derivative with respect to the spacetime metric $g_{\mu\nu}$. The tensor $K_{\alpha\beta\mu\nu}$ is defined by
\begin{equation}\label{K}
\begin{aligned}
K_{\alpha\beta\mu\nu}=&\frac{\partial P}{\partial R^{\alpha\beta\mu\nu}}\\
=&12(\tfrac{1}{2}{R_{\a\b}}^{\r\s}R_{\m\n\r\s}+6{{{R_{\a}}^{\r}}_{[\m}}^{\s}R_{\n]\s\b\r}+2g_{\b[\m}R_{\n]\s\a\r}R^{\r\s}\\
&-2g_{\a[\m}R_{\n]\s\b\r}R^{\r\s}-4R_{\r[\m}g_{\n][\a}{R_{\b]}}^{\r}-2R_{\a[\m}R_{\n]\b}).
\end{aligned}
\end{equation}
The tensor $H_{\mu\nu}$ represents the contribution from the cubic term $P$. Observing the form of  $H_{\m\n}$  we
mention that the complication and length of the above general field equations does not easily allow us to perform the Hamiltonian analysis and check whether they present unhealthy high derivatives, in the
sense that they do not correspond to healthy extra degrees of freedom. Nevertheless, this
necessary check, i.e. ensuring the condition (iii) above, can be performed around 
the cosmological
background on which we focus in this work, as we will shortly see.

In order to proceed to the cosmological application of cubic gravity we consider a   flat
homogeneous and isotropic FLRW geometry with
  metric
\begin{equation}
\label{FRWmetric}
ds^{2}=-dt^{2}+a^{2}(t)\delta_{ij}dx^{i}dx^{j}\,,
\end{equation}
where $a(t)$ is the scale factor. Additionally, as usual we consider the matter
Lagrangian $\Lm$ to correspond to a  perfect fluid with energy-momentum tensor
$
T_{\m\n}=(\rho_m+p_m)u_\m u_\n+p_m g_{\m\n},
$
where $\rho_m$ and $p_m$ are   the energy density and pressure respectively and $u_\m$
is the fluid four-velocity. Under these considerations the general field equations
\eqref{eom} give rise to the Friedmann equations. These equations can now be easily proven 
to be second order if we make the additional parameter reduction
  \begin{eqnarray}
 \label{prop22}
 \b_6=4\b_2+2\b_3+8\b_4+\b_5.
\end{eqnarray}
 In particular, they are written as
\begin{eqnarray}
\label{Fr1a}
3H^2&=&\k\left( \r_m+6 \a \tilde{\b} H^6\right)+\Lambda,
\\
3H^2+2\dot{H}&=&-\k\left[ p_m-6\a \tilde{\b}  H^4( H^2+2  \hd)\right]+\Lambda,
\label{Fr2a}
\end{eqnarray}
where $H=\frac{\dot a}{a}$ is the Hubble parameter, with dots
denoting derivatives with respect to $t$, and where we have defined the parameter
  \begin{eqnarray}
\tilde{\b}\equiv -\b_1+4\b_2+2\b_3+8\b_4.
\end{eqnarray}
Finally, note that in FLRW geometry, and under the parameter relations
(\ref{prop}) and (\ref{prop22}), the cubic invariant $P$ acquires the simple form
  \begin{eqnarray}
  \label{Prel}
P=6\tilde{\beta} H^4(2H^2+3\dot{H}),
\end{eqnarray}
which includes only up to first-order derivatives and hence the Friedmann equations are
up to second order, as indeed they are constructed to be.

Observing the form of the two Friedmann equations we deduce that they can be rewritten
in the usual way
\begin{eqnarray}
\label{Fr1}
3H^2&=&\k\left( \r_m+ \r_{cub}\right),
\\
3H^2+2\dot{H}&=&-\k\left( p_m +p_{cub}\right),
\label{Fr2}
\end{eqnarray}
where
\begin{eqnarray}
\label{rholinear}
&&
\r_{cub} \equiv    6\b H^6+\frac{\Lambda}{\k},\\
&&
p_{cub}\equiv  -6  \b  H^4( H^2+2  \hd)-\frac{\Lambda}{\k},
\label{plinear}
\end{eqnarray}
and where we have redefined the sole parameter as $\b\equiv\alpha\tilde{\b}$.
Hence, in cubic gravity we obtain an effective sector that incorporates the effects of
the cubic modification of the action. Finally, one can immediately see that this
effective sector is conserved, namely
\begin{eqnarray}
\dot{\r}_{cub}+3H(\r_{cub}+p_{cub})=0,
\end{eqnarray}
while the matter sector is conserved too, namely
\begin{eqnarray}
\dot{\r}_m+3H(\r_m+p_m)=0.
\end{eqnarray}

 We close this section by constructing $f(P)$ gravity. In the above we described how one
can construct a subclass of cubic gravity which in four dimensions is nontopological
and in a cosmological spacetime it leads to second-order field equations. The theory is
based on the cubic invariant $P$ of (\ref{P}), under the parameter choices
(\ref{prop}) and (\ref{prop22}) which ensure the aforementioned requirements. 
Nevertheless,
having such an invariant we may generalize cubic gravity to $f(P)$ gravity, characterized
by the action
\begin{equation}
S=\int\sqrt{-g}d^4x\left[\frac{R}{2\kappa}+f(P)\right],
\label{actionfP}
\end{equation}
where  $f(P)$ is an
arbitrary function of   $P$.
Variation of the action $S+S_m$ leads
to the field equations
\begin{equation}\label{eom2}
G_{\mu\nu}=\kappa (T_{\m\n}+\tilde{H}_{\m\n}),
\end{equation}
with  $T_{\m\n}$ still given by (\ref{Tmn}) and
\begin{equation}\label{eqf(p)}
\begin{aligned}
\tilde{H}_{\m\n}&=-\frac{2}{\sqrt{-g}}\frac{\delta (\sqrt{-g}f(P))}{\delta
g^{\mu\nu}},\\
&=g_{\mu\nu}f(P)+{R^{\a\b\r}}_{(\mu}\tilde{K}_{\nu)\r\a\b}+2\nabla^{\a}\nabla^{\b}\tilde{K}_{\a(\m\n)\b}.
\end{aligned}
\end{equation}
Here, tensor $\tilde{K}_{\a\b\m\n}$ is given in terms of tensor \eqref{K} as $\tilde{K}_{\a\b\m\n}=f'(P)K_{\a\b\m\n}$ with primes denoting the derivative of a function with respect to its argument.

In the case of FLRW geometry, the field
equations (\ref{eom2}) provide the two Friedmann equations, namely
 \begin{eqnarray}
 \label{Fr1fP}
3H^2&=&\k( \r_m+ \r_{f_P}),\\
3H^2+2\dot{H}&=&-\k(p_m+ p_{f_P}),
 \label{Fr2fP}
\end{eqnarray}
where
\begin{eqnarray}
\label{rhofP}
&&\r_{f_P}\equiv-f(P)-18\tilde{\b} H^4(H\partial_t-H^2-\dot{H})f'(P),\\
&&p_{f_P}\equiv f(P)+6\tilde{\b}
H^3\left[H\partial_t^2+2(H^2+2\hd)\partial_t-3H^3-5H\hd\right]f'(P),
\label{pfP}
\end{eqnarray}
and $\partial_t\equiv \frac{\partial}{\partial t}$ and $\partial^2_t\equiv
\frac{\partial^2}{\partial t^2}$. Additionally, note that both the matter and the
effective $f(P)$ sector are conserved, namely
\begin{eqnarray}
\dot{\r}_m+3H(\r_m+p_m)&=&0,\\
\dot{\r}_{f_P}+3H(\r_{f_P}+p_{f_P})&=&0.
\end{eqnarray}
Lastly, in the case where $f(P)=\alpha P-\frac{\Lambda}{\k}$, $f(P)$ gravity gives the
simple cubic gravity of (\ref{action0}).

The action \eqref{actionfP} can be recast into a (classically) dynamically equivalent action where the system of equations \eqref{Fr1fP} and \eqref{Fr2fP} can be put explicitly as a second-order system by introducing two scalar variables related by $\varphi=f(\phi)$. The details for this procedure can be found in the Appendix.

\section{Cosmological applications}
\label{CosmAppl}

In this section we investigate the cosmological applications of cubic and $f(P)$ gravity
at both early and late times. 
Concerning the early-time application as usual we may neglect the matter sector. We are
interested in obtaining the de Sitter solution, which is the basis of the inflation
realization. Let us consider first the simple cubic gravity, namely the first
Friedmann equation (\ref{Fr1}) without the matter term. In this case we can easily see
that one obtains the de Sitter solution
\begin{eqnarray}
H^2=\frac{6^{1/3}\k\beta+\zeta^{2/3}}{
-6^{\frac{2}{3}}\k\beta\zeta^{
1/3}}=const.,
\end{eqnarray}
where $\zeta=3\k^2\b^2\Lambda+\sqrt{3}\sqrt{\k^3\b^3(3\k\b\Lambda^2-2)}$, provided $\b<0$. 
When the 
cosmological constant is absent we get that $H^2=(2\kappa\beta)^{-1/2}$ for $\b>0$. In 
similar
lines, in the case of $f(P)$ gravity, i.e. of the Friedmann equation (\ref{Fr1fP}),
without the matter sector, we can see that in general we can obtain the de Sitter
solution 
\begin{equation}
H^2=\frac{\kappa}{3}\left[18\tilde{\b}H^6f'(P)-f(P)\right]=const.
\end{equation}
and $\dot{H}=0$. As we observe, in both theories the de Sitter
solution is driven by an effective cosmological constant that is constructed purely from
the cubic terms, even if an explicit cosmological constant is absent.

Let us now focus on the late-time evolution.
We first examine the basic cubic gravity,
namely (\ref{action0}), where the correction in the general relativity action is just
$P$. In this case the cubic correction constitutes the effective dark-energy sector,
whose energy density and pressure are
\begin{eqnarray}
\rho_{DE}&\equiv&\r_{cub}\\
p_{DE}&\equiv& p_{cub},
\end{eqnarray}
with $\r_{cub}$ and $p_{cub}$ given respectively by (\ref{rholinear}) and
(\ref{plinear}),
while the effective dark-energy equation-of-state parameter is
\begin{eqnarray}
w_{DE}\equiv  \frac{p_{DE}}{\rho_{DE}}.
\label{wDEdef}
\end{eqnarray}
In order to proceed we define the density parameters $\Omega_i\equiv
\kappa\rho_i/(3H^2)$, with $i$ standing for matter and dark energy. Lastly,
it proves convenient to introduce the deceleration
parameter $q$, which reads as
  \begin{equation}
  \label{qdeccel}
  q\equiv-1-\frac{\dot{H}}{H^2}=\frac{1}{2}+\frac{3}{2}\left(w_m\Omega_m+w_{DE}\Omega_{DE}
  \right).
\end{equation}

\begin{figure}[ht]
\includegraphics[scale=0.45]{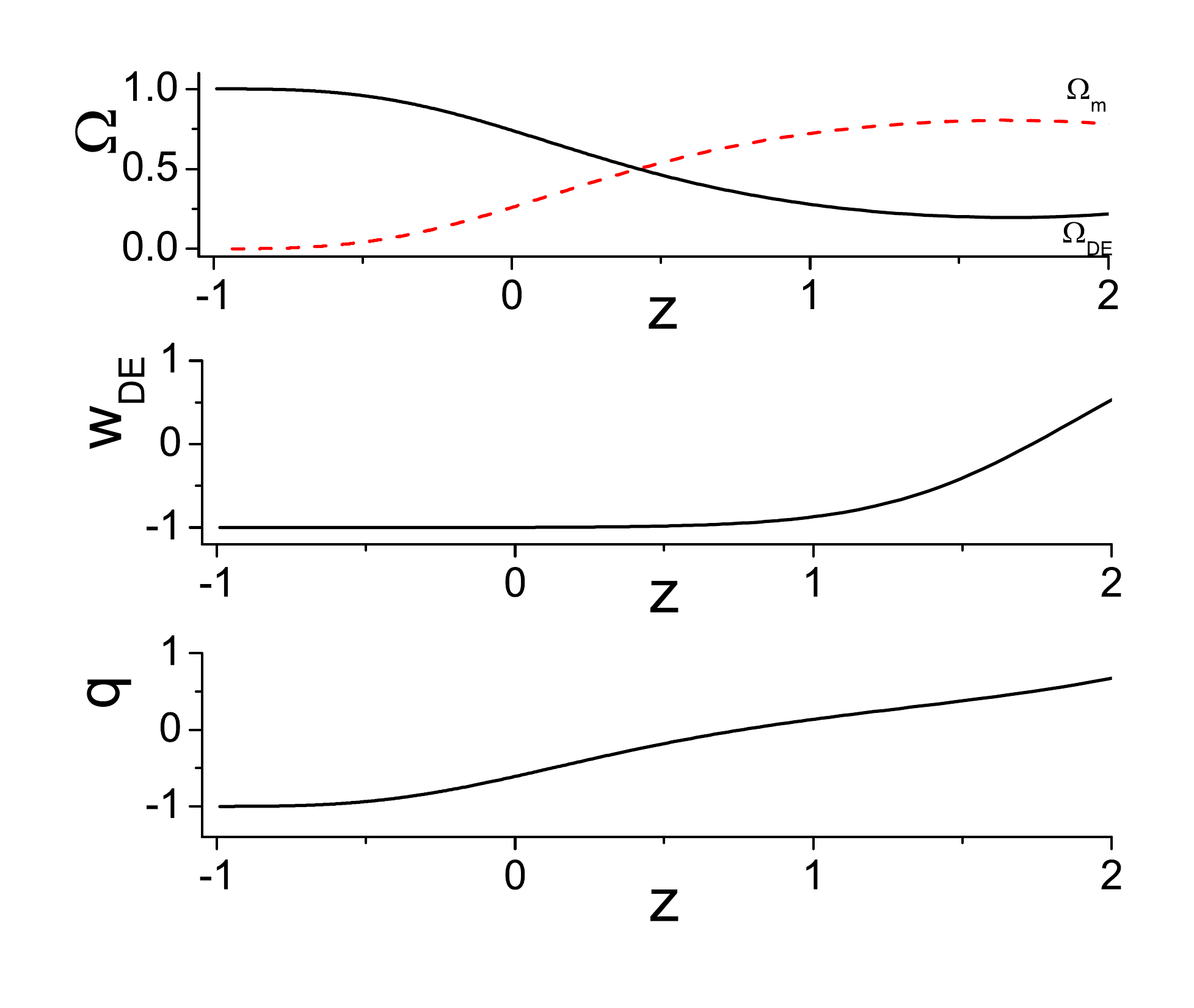}
\caption{
{\it{ Upper graph: The evolution of the effective dark energy
density parameter $\Omega_{DE}$ (black-solid)  and of the matter density
parameter $\Omega_{m}$ (red-dashed), as a function of the redshift $z$,
in the case of simple cubic gravity (\ref{action0}), for  $\beta=0.005$ and $\Lambda=1$,
in units where $\kappa=1$. We have imposed the initial conditions
$\Omega_{DE}(z=0)\equiv\Omega_{DE0}\approx0.69$  in agreement with observations.
  Middle graph: The evolution of the corresponding effective dark
energy equation-of-state parameter $w_{DE}$ from (\ref{wDEdef}). Lower graph:  The
evolution of the
corresponding   deceleration parameter $q$ from  (\ref{qdeccel}).
}} }
\label{LinearP}
\end{figure}

We elaborate the Friedmann equations  (\ref{Fr1}) and (\ref{Fr2}) numerically, using the
redshift $z$ as the independent variable, defined  as  $
 1+z=a_0/a$ with the present scale factor set to $a_0=1$. Concerning the initial 
conditions, we set
$\Omega_{DE}(z=0)\equiv\Omega_{DE0}\approx0.69$  and thus
$\Omega_m(z=0)\equiv\Omega_{m0}\approx0.31$  as required by observations
\cite{Ade:2015xua}. Additionally, we consider the matter sector to be dust, namely we set
 $w_m=0$.  In the upper graph of Fig. \ref{LinearP} we depict
$\Omega_{DE}(z)$ and $\Omega_{m}(z)=1-\Omega_{DE}(z)$. In the middle graph we present the
corresponding behavior of $w_{DE}(z)$ given by  (\ref{wDEdef}). Finally, in the
lower graph we show the deceleration parameter as it arises from  (\ref{qdeccel}).

From the upper graph of Fig. \ref{LinearP} we observe that we can obtain the usual
thermal history, that is the sequence of matter and dark-energy eras, while in the future
($z\rightarrow-1$) the Universe asymptotically results in a complete dark-energy
dominated, de Sitter phase. Moreover,  from
the third graph of Fig. \ref{LinearP} we see that the transition
from deceleration to acceleration is realized at $z\approx 0.5$  in agreement with the
observed behavior. Additionally, from the middle graph of Fig. \ref{LinearP} we observe
that the effective dark-energy equation-of-state parameter $w_{DE}$ quickly acquires the value $-1$.

Nevertheless, we must mention here that in the simple cubic gravity  (\ref{action0}) the
late-time acceleration is triggered mainly by the explicit cosmological constant term
$\Lambda$,
while the cubic term, quantified by the parameter $\beta$ in
(\ref{rholinear}) and (\ref{plinear}), plays a secondary role (a similar result was found 
in a different context in \cite{Edelstein01}). The reason is
that
since the cubic correction in (\ref{rholinear}) is proportional to $H^6$, if this term is
to play the main role in the late-time acceleration then at earlier times it will be
large enough in order to spoil the desired thermal history of the Universe and lead to an
``early dark energy'' \cite{Bartelmann:2005fc} behavior. However, even in the case where
the cubic term plays a
secondary role in the late-time acceleration of the Universe, its effect can still be
significant at intermediate times (see the middle graph of  Fig. \ref{LinearP} where the
deviation from $\Lambda$CDM scenario is clear) as well as at the perturbation level
reflected in the matter clustering and the large scale structure (LSS). The detailed
analysis of the perturbations and the comparison with LSS data such as the $f\sigma_8$
ones will be handled in a separate project.

Let us now investigate the extended cubic gravity, namely the $f(P)$ gravity
(\ref{actionfP}). In this case, using (\ref{rhofP}) and (\ref{pfP}), the effective dark-energy density and pressure are given by
 \begin{eqnarray}
\rho_{DE}&\equiv&\r_{f_P}\\
p_{DE}&\equiv&p_{f_P},
\end{eqnarray}
and the effective dark-energy equation-of-state parameter as usual is
$w_{DE}\equiv  p_{DE}/ \rho_{DE}$. As a first example we consider the case where
 \begin{eqnarray}
f(P)=\alpha\sqrt{P},
\label{sqrtfP}
\end{eqnarray}
without considering an explicit cosmological constant. Additionally, we merger the
coupling parameter $\alpha$ with the parameter $\tilde{\beta}$ of $P$ in (\ref{Prel})
through $\b\equiv\alpha\tilde{\b}$.

 \begin{figure}[ht]
\includegraphics[scale=0.45]{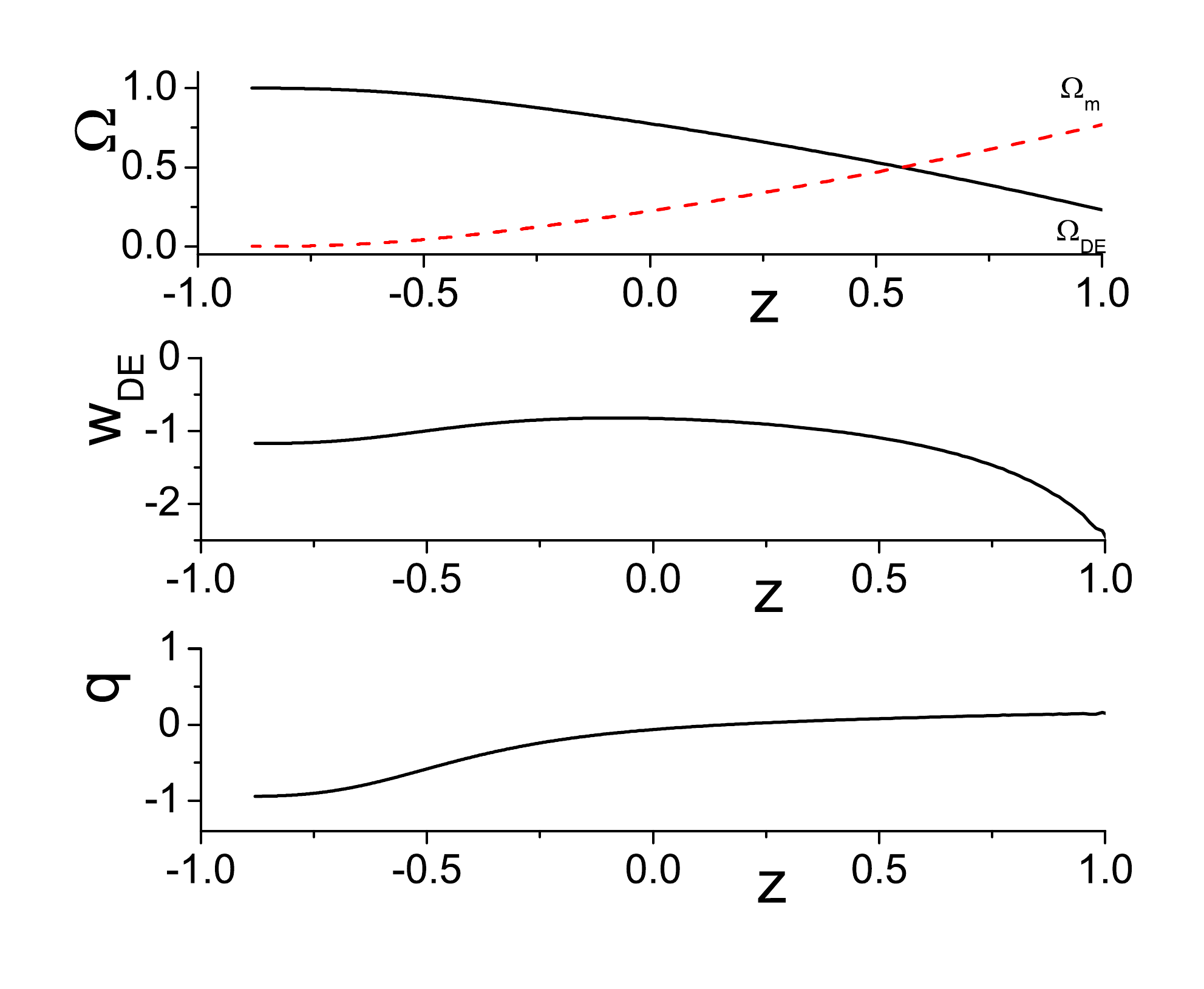}
\caption{
{\it{ Upper graph: The evolution of the effective dark-energy
density parameter $\Omega_{DE}$ (black-solid)  and of the matter density
parameter $\Omega_{m}$ (red-dashed), as a function of the redshift $z$,
in the case of $f(P)$ gravity (\ref{actionfP}), under the square choice (\ref{sqrtfP}),
for  $\beta=5$ in units where $\kappa=1$. We have imposed the initial conditions
$\Omega_{DE}(z=0)\equiv\Omega_{DE0}\approx0.69$  in agreement with observations.
  Middle graph: The evolution of the corresponding effective dark
energy equation-of-state parameter $w_{DE}$. Lower graph:  The
evolution of the
corresponding   deceleration parameter $q$.
}} }
\label{sqrtP}
\end{figure}
 \begin{figure}[ht]
\includegraphics[scale=0.45]{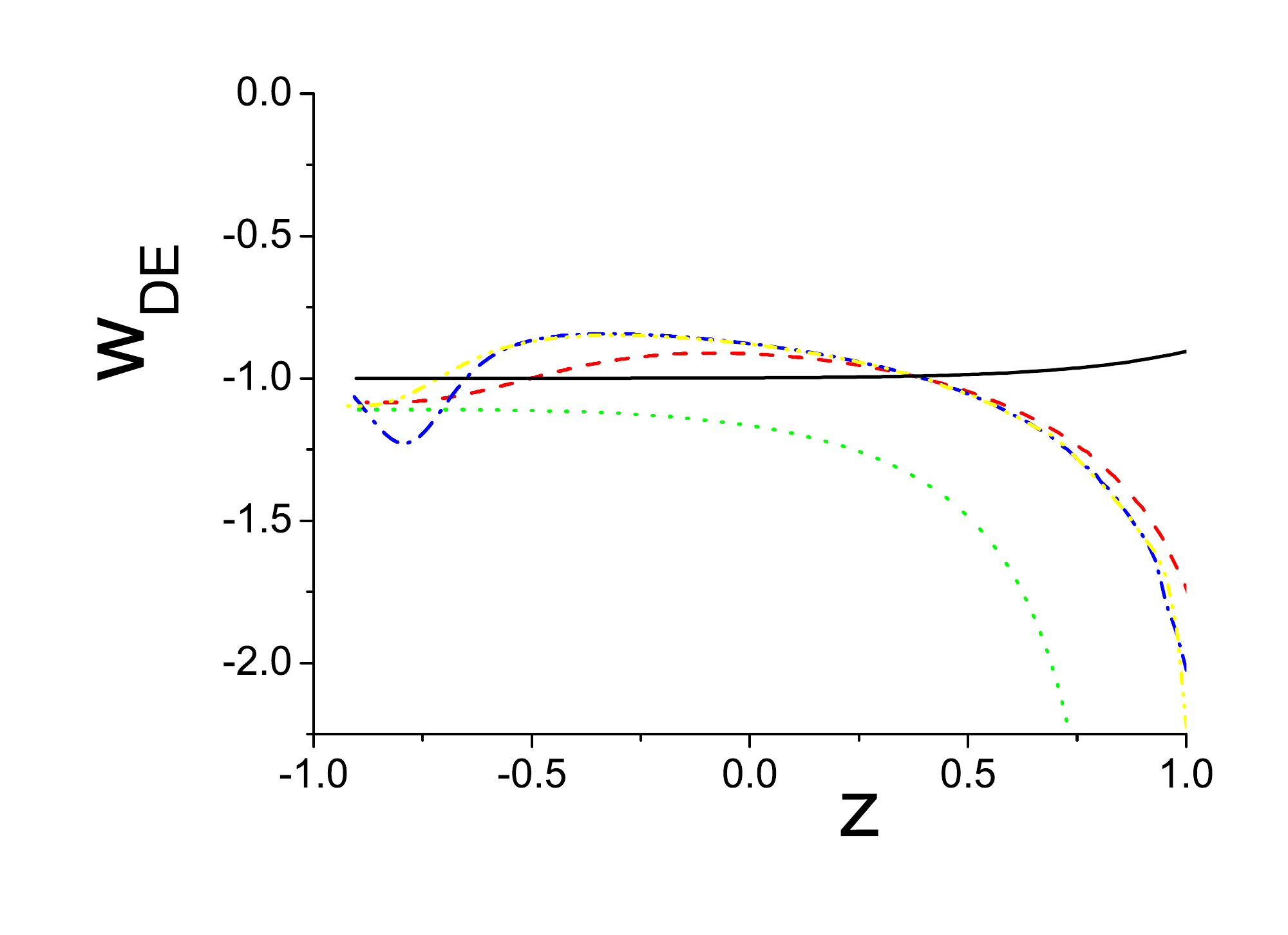}
\caption{
{\it{  The evolution of the equation-of-state parameter  $w_{DE}$  of the effective dark
energy of $f(P)$ gravity, for various forms of $f(P)$: $f(P)=\beta P-\frac{\Lambda}{\k}$
with $\b\equiv\alpha\tilde{\b}=0.005$ and $\Lambda=1$ (black - solid),
$f(P)=\alpha\sqrt{P}$
with $\b\equiv\alpha\tilde{\b}=5$ (red - dashed), $f(P)=\alpha P^{1/3}$
with $\b\equiv\alpha\tilde{\b}=-10$ (green - dashed), $f(P)=\alpha\sqrt{P}+\gamma P$
with $\b\equiv\alpha\tilde{\b}=1$  and $\zeta\equiv\gamma\tilde{\b}=0.1$  (blue dashed-dotted), and  $f(P)=\alpha P^{1/3}+\gamma P$
with $\b\equiv\alpha\tilde{\b}=5$  and $\zeta\equiv\gamma\tilde{\b}=0.050$  (yellow dashed-dotted).
We have imposed the initial conditions
$\Omega_{DE}(z=0)\equiv\Omega_{DE0}\approx0.69$ and we use units where $\kappa=1$.
}} }
\label{wDE}
\end{figure}

We elaborate the Friedmann equations
(\ref{Fr1fP}) and (\ref{Fr2fP}) numerically, considering the matter sector to be dust. In
the
upper graph of Fig. \ref{sqrtP} we present
$\Omega_{DE}(z)$ and $\Omega_{m}(z)$, in the middle graph we draw $w_{DE}(z)$, and in the
lower graph we depict the deceleration parameter $q(z)$. As we observe, once again we
obtain  the sequence of matter and dark-energy eras and the onset of late-time
acceleration. However, the significant advantage compared to the simple cubic gravity of
Fig. \ref{LinearP} is that now the above behavior is obtained although we have not
considered an explicit cosmological constant; namely it arises purely from the $f(P)$
modification. This can also be seen in the evolution of $w_{DE}$, which exhibits a
dynamical behavior different than $\Lambda$CDM at all times, with an asymptotic value
slightly into the phantom regime.

In order to examine the effect of $f(P)$ forms on  $w_{DE}$, in  Fig. \ref{wDE} we
present the evolution of  $w_{DE}(z)$ for various $f(P)$ choices. In particular, apart
from the simple cubic gravity of  Fig. \ref{LinearP}, which is obtained for  $f(P)=\beta
P-\frac{\Lambda}{\k}$, and for the square-root $f(P)$ gravity of (\ref{sqrtfP}) of Fig.
\ref{sqrtP}, we consider the cubic-root case $f(P)=\alpha P^{1/3}$, as well as the
combined cases  $f(P)=\alpha\sqrt{P}+\gamma P$ and  $f(P)=\alpha P^{1/3}+\gamma P$.

As we observe, one can fix the model parameters in order to obtain a $w_{DE}$  at
present ($z=0$) close to the cosmological constant value, in agreement with observations,
while the past and future behavior can be different for different scenarios. As we
mentioned earlier, we note that apart from the linear case, such behavior is achieved
without the need for an explicit cosmological constant, which is a significant advantage.
Furthermore, note that $w_{DE}$ can be  lying in the quintessence
regime or in the phantom regime or it can experience the phantom-divide crossing during the
cosmological evolution, which reveals the capabilities of the theory at hand, since it is
known that the phantom regime cannot be easily obtained \cite{Nojiri:2013ru}.

\section{Conclusions}
\label{Conclusions}

In this work we constructed cubic gravity and its extension and we investigated their
cosmological applications for both the early- and late-time Universe. Cubic gravity is based on
the invariant $P$ that is constructed using particular cubic contractions of the
Riemann tensor, such that the theory (i) possesses a spectrum identical to that of
Einstein gravity, i.e. the metric perturbation on a maximally symmetric background
propagate only a transverse and massless graviton;
(ii) it is neither topological nor trivial in four dimensions, and  (iii) it is 
defined such 
that it is independent of the dimensions.
By relaxing this last condition, and tuning the coupling parameters, the theory admits 
second-order 
field equations for a four-dimensional cosmological ansatz without the inclusion of any 
additional 
term. These features make  cubic gravity a
well-behaved candidate for modified gravity. Additionally, apart from simple cubic
gravity, which uses $P$ in the Lagrangian, we constructed extensions, namely $f(P)$
gravity.

Applying both simple cubic and $f(P)$ gravity in a cosmological framework,  we saw that
at early times one can easily obtain inflationary, de Sitter solutions,
which are  driven by an effective cosmological constant that is constructed purely from
the cubic terms, even if an explicit cosmological constant is absent.

Concerning the late-time evolution, we showed that the cubic terms construct an
effective dark-energy sector of gravitational origin. In the case of simple cubic gravity
we obtained the usual thermal history, i.e. the sequence of matter and dark-energy eras,
with the onset of late-time acceleration at around $z\approx 0.5$  in agreement with
observations, while in the future the Universe asymptotically results in a completely
dark-energy dominated, de Sitter phase. Nevertheless, if we desire not to spoil the
Universe's evolution at early times then we find that  the background
late-time acceleration is triggered mainly by the explicit cosmological constant, with
the cubic term playing a secondary role (which could however be significant at the
perturbation level).

However, when we proceeded to $f(P)$ gravity we found that we can obtain the    usual
thermal history and the onset of late-time acceleration even without considering an
explicit cosmological constant, namely as a result of  the $f(P)$ corrections purely.
Moreover, examining the evolution of the effective dark-energy equation-of-state
parameter $w_{DE}$ we showed that its behavior is determined by the $f(P)$ form as well
as the parameter choices, and it can be quintessencelike or phantomlike or it can experience
the phantom-divide crossing during the evolution. Thlikeese features reveal the capabilities
of $f(P)$ gravity and offer a good motivation for its further investigation.

Finally, we mention  that there are additional studies that
need to be  performed before  cubic and $f(P)$ gravity can be
considered as a successful candidate for the description of nature. Firstly, one should
use data from type Ia supernovae (SNIa),
cosmic microwave background (CMB) shift parameter, baryon acoustic oscillations (BAO), and
direct Hubble constant observations, in order to extract constraints on the involved
forms and parameters. Furthermore, one needs to perform a detailed perturbation
analysis and confront the theory with data from   CMB temperature and polarization as
well as from LSS (such as the $\sigma_8$ ones). Moreover, one
could  perform a phase-space analysis in order to reveal the
global behavior of $f(P)$ cosmology. Lastly, going beyond the cosmological framework, as 
we mentioned above  the original ECG theory can be corrected in order to admit black hole 
and cosmological solutions; however these solutions coexist in a disjoint set of values 
for the cubic coupling parameter. It would be interesting to investigate whether the above 
$f(P)$ modification provides enough 
freedom to obtain black hole solutions in a common range of parameters.
These necessary investigations lie beyond the scope
of this work and are left for future projects.

\section*{Acknowledgments}
This article is based upon work from COST Action ``Cosmology and Astrophysics Network for
Theoretical Advances and Training Actions,'' supported by the European Cooperation in
Science and Technology (COST). C.E. acknowledges financial support given by Becas Chile, 
Comisi\'on Nacional de Investigaci\'on Cient\'ifica y Tecnol\'ogica.

\newpage
\appendix*
\section{Equations of motion in $f(P)$ gravity as second-order field equations}
\label{appendix}

The $f(P)$ gravity can be recast into a (classically) dynamically equivalent action where the Friedmann equations \eqref{Fr1fP} and \eqref{Fr2fP} can be put explicitly as a second-order system. For this let us consider the following action endowed with scalar variables $\varphi$ and $\phi$:
\begin{equation}
S=\int\sqrt{-g}d^4x\left[\frac{R}{2\kappa}+f(\phi)+\varphi(P-\phi)\right].
\label{actionBD}
\end{equation}
Variation with respect to $\varphi$ yields $\phi=P$. Replacing this algebraic constraint back into action \eqref{actionBD} yields the action \eqref{actionfP}. Considering the variation with respect to $\phi$, it is obtained that $\varphi=f'(\phi)$. Variation with respect to the metric $g_{\mu\nu}$ leads to
\begin{equation}
\begin{aligned}
G_{\mu\nu}&=\kappa(T_{\mu\nu}+\tilde{H}_{\m\n})\\
\tilde{H}_{\m\n}&\equiv g_{\mu\nu}f(\phi)+\varphi{R^{\a\b\r}}_{(\mu}K_{\nu)\r\a\b}+2\nabla^{\a}\nabla^{\b}\left[\varphi K_{\a(\m\n)\b}\right],
\end{aligned}
\end{equation}
which retrieves equations equivalent to \eqref{eqf(p)} when the algebraic constraints are considered. Then, the Friedmann equations \eqref{Fr1fP} and \eqref{Fr2fP}, take the same form but with an effective sector given by
\begin{eqnarray}
&&\r_{f_P}\equiv-f(\phi)-18\tilde{\b} H^4(H\partial_t-H^2-\dot{H})\varphi,\\
&&p_{f_P}\equiv f(\phi)+6\tilde{\b}
H^3\left[H\partial_t^2+2(H^2+2\hd)\partial_t-3H^3-5H\hd\right]\varphi,
\end{eqnarray}
which clearly possesses second-order field equations.

\end{document}